\documentclass[a4paper,1za1pt]{article}
\pdfoutput=1 

\usepackage{jcappub} 

\usepackage{subfigure}
\usepackage{xcolor}

\usepackage[english]{babel}
\usepackage[utf8x]{inputenc}
\usepackage[T1]{fontenc}

\newcommand{\rmax}{r_{\rm{max}}}

\newcommand{\mpch}{\rm{Mpc}/h}
\newcommand{\xidos}{\xi^{(2)}}

\usepackage{xcolor}
\definecolor{light-gray}{gray}{0.95}
\newcommand{\code}[1]{\colorbox{light-gray}{\texttt{#1}}}

\makeatletter
\gdef\@fpheader{}
\makeatother

\begin{document}
    
\title{Lya2pcf: an efficient pipeline to estimate two- and three-point correlation functions of the Lyman-$\alpha$ forest}

\author[1,2]{Josue De-Santiago,}
\author[1]{Rafael Gutiérrez-Balboa,}
\author[3]{Gustavo Niz}
\author[3]{and Alma X. González-Morales}

\affiliation[1]{Depto.  de Física, Centro de Investigación y de Estudios Avanzados del IPN, A.P. 14-740, 07000 Ciudad de México, México.}
\affiliation[2]{Secretaría de Ciencia, Humanidades, Tecnología e Innovación,  Av.   Insurgentes  Sur  1582,  Colonia  Crédito Constructor, Del.  Benito Juárez, 03940, Ciudad de México, México.}
\affiliation[3]{Departamento de Física, DCI, Campus León, Universidad de Guanajuato, Loma del Bosque 103, León, Guanajuato C.~P.~37150, México.}

\emailAdd{Josue.desantiago@cinvestav.mx}
\emailAdd{rafael.gutierrez@cinvestav.mx}
\emailAdd{g.niz@ugto.mx}
\emailAdd{gonzalez.alma@ugto.mx}


\abstract{ 
Studying the matter distribution in the universe through the Lyman-$\alpha$ forest allows us to constrain small-scale physics in the high-redshift regime. Spectroscopic quasar surveys are generating increasingly large datasets that require efficient algorithms to compute correlation functions. Moreover, cosmological analyses based on Lyman-$\alpha$ forests can significantly benefit from incorporating higher-order statistics alongside traditional two-point correlations.
In this work, we present Lya2pcf, a pipeline designed to compute three-dimensional two-point and three-point correlation functions using Lyman-$\alpha$ forest data. The code implements standard algorithms widely used in current spectroscopic surveys for computing the two-point correlation function with its distortion matrix,  covariance matrices; and it naturally extends the two-point estimator to three-point correlations. Thanks to GPU optimization, Lya2pcf achieves a substantial reduction in computational time for both the two-point correlation function and its distortion matrix when compared to the widely used PICCA code.
We apply Lya2pcf to data from the Sloan Digital Sky Survey (SDSS) sixteenth data release (DR16) and a Dark Energy Spectroscopic Instrument Year-5 (DESI Y5) mock dataset, demonstrating overall performance gains over PICCA, especially on GPUs. We show the first measurement of the anisotropic three-point correlation function on a large spectroscopic sample for all possible triangles with scales up to 80 Mpc/h. The estimator's fast computation and the resulting signal-to-noise ratio ---above one for many triangle configurations--- demonstrate the viability of incorporating three-point statistics into future cosmological inference analyses, particularly with the larger datasets expected from Stage IV spectroscopic surveys.
}
\maketitle


\section{Introduction}
The Lyman-$\alpha$ (Ly-$\alpha$) forests are structures observed in the spectrum of distant quasars (QSOs). As light travels to us it encounters neutral hydrogen clouds that
absorb sections of the quasar spectrum. These absorption features are located at different wavelengths because of the redshift of light due to the expansion of the universe, creating a \textit{forest} of absorption lines in the observed QSO spectrum. 
These absorption features trace the underlying matter density field on light-ray trajectories, allowing us to probe models of the matter distribution and evolution in the universe at redshifts $>2.1$, larger than other tracers such as galaxies \cite{Rauch:1998xn, McQuinn:2015icp}. Moreover, the data resolution of such Ly-$\alpha$ tracers allows for studies of small-scale physics, which are relevant to understand neutrino masses, dark matter properties, test gravitational interactions, physics of the intergalactic medium, among other physical phenomena.
 
A fundamental statistical tool to describe the distribution of matter and derive cosmological constraints is to compute pair correlations of its tracers. For the particular case of Lyman-$\alpha$ forests, studies have focused mainly on correlations along the line of sight in Fourier space \cite{croft1999power,SDSS:2004kjl,McDonald:1999dt,BOSS:2013rpr,2015JCAP11011,eBOSS:2018qyj,Karacayli:2023afs,Karacayli:2025svi,Ravoux:2025uik} and on three dimensions in real space \cite{slosar2011lyman, Busca_2013,Agathe:2019vsu,Blomqvist:2019rah,2025JCAP01124,DESI:2025zpo}. We pay particular attention to the 3D real space two-point correlations, where robust signals have been obtained thanks to the development of surveys focused to measure a large quantity of quasars and their corresponding forests, like the Baryon Oscillation Spectroscopic Survey (BOSS). Both BOSS and its continuation extended BOSS (eBOSS) steadily increased the number of measured Lyman-$\alpha$ forests from $~14,000$ \cite{slosar2011lyman} up to $>200,000$ during the final years of operation \cite{BOSS:2017fdr,Agathe:2019vsu,Blomqvist:2019rah,des2020completed}. In recent years, the Dark Energy Spectroscopic Instrument (DESI) \cite{DESI:2022xcl} has  increased the number of measured quasar spectra by an order of magnitude with respect to eBOSS, significantly decreasing the statistical errors of cosmological estimations. DESI measured $88,000$ forests in the Early Data Release \cite{2024AJ16858}, $420,000$ in the first data release (DESI DR1) \cite{2025JCAP01124}, made public earlier this year, and provided the first analysis of the second data release (DESI DR2) using $820,000$ forests \cite{DESI:2025zpo}. This increase in the number of measured forests advocates for efficient computational tools that can analyze its statistical properties with precision and at reasonable computational costs.

A Gaussian random field can be fully characterized by its two-point correlation function. However, if there are initial non-Gaussianities or the evolution of initial Gaussian data is driven by non-linear equations, the resulting distribution has non-trivial higher (than two) order correlation functions. This is the case of the matter evolution in our Universe. Therefore, using higher-order statistics alongside their two-point counterparts would help to break parameter degeneracies, better understand systematics, explore new physics (e.g., modified gravity or parity violation), and constrain primordial non-Gaussianities. It is important to stress that there are not that many studies on the three- or higher-point statistics using the Lyman-$\alpha$ forest (some of the few examples are \cite{Mandelbaum, Zaldarriaga:2000rg, Viel:2003fp,Tie:2019tpi, Maitra:2019trn, Maitra:2020jrk, Adari:2024vkf, Karacayli:2024udo, arXiv:241009150}).

In the case of the Ly-$\alpha$ forest, there is a popular set of tools, under the name of PICCA\footnote{\url{https://github.com/igmhub/picca}}, to calculate the auto-correlation of the forest and its cross-correlation with quasars. The software runs on Python and is the standard approach in DESI \cite{DESI:2022xcl}.  The purpose of this work is to develop an alternative infrastructure for the Lyman-$\alpha$ forest analysis, with high-level optimization, including the use of GPU computation, to accommodate for independent calculations of the two-point correlations and also to incorporate efficient algorithms of three-point statistics, which should help constrain our understanding of the universe.

The paper is organized as follows. In Section \ref{2} we review the theory of correlation functions and write the estimators for the two and three point correlation functions that will be used in our pipeline. In Section \ref{3} we review technical aspects of our pipeline and measure performance metrics. Finally, in Section \ref{4} we discuss our results and conclude.

\section{Correlation functions with the Lyman-$\alpha$ forest}\label{2}
The modern approach to structure formation in the universe as a stochastic process benefits from the fundamental objects known as correlation functions. By definition, the $n$-point correlation function is
\begin{equation}
  \xi^{(n)}(\boldsymbol{r_1},\boldsymbol{r_2}\ldots,\boldsymbol{r_{n-1}}) = \left\langle \delta(\boldsymbol{x})\delta(\boldsymbol{x}+\boldsymbol{r_1})\ldots \delta(\boldsymbol{x}+\boldsymbol{r_{n-1}})\right\rangle \,,
  \label{def_correlation}
\end{equation}
where the $\delta$ field corresponds to the inhomogeneous fluctuation in the quantity of interest ---energy density, number density, etc.--- and $\langle...\rangle$ denotes the expectation value over the sample in question. Statistical homogeneity implies the correlation function depends on $n-1$ displacement vectors, $\boldsymbol{r_i}$, among the sampled data points.
In the two-point correlation function (2PCF) $\xidos(\boldsymbol{r})$, the statistical isotropy reduces the function dependence to only
the magnitude of the separation between each pair of points, $\xidos(r)$. However, given that spectroscopic data lives in the anisotropic redshift space, a second variable is needed to characterize the anistropic correlation function, for example, the angle of these pair separation with respect to the line of sight. Ly-$\alpha$ analysis often uses instead the parallel and
perpendicular directions to the line of sight, $\xidos(r_\parallel,r_\bot)$.

Fluctuations in the Lyman-$\alpha$ forest resemble those of the CMB temperature anisotropies, though they have support on a volume instead of the last scattering surface. There are two approaches to calculate correlation functions in the CMB, either one writes a likelihood and maximises it (using, for example, a Newton-Rapson method) or proposes a particular form and checks that the estimator is unbiased and optimal (e.g. \cite{Bond:1998zw, Seljak:1997wx})\footnote{Examples of the two approaches in the Ly-$\alpha$ 1D power spectrum can be found in recent DESI studies of \cite{2023MNRAS526,Karacayli:2023afs,Ravoux:2025uik,Karacayli:2025svi}.}. We will focus on the second approach for the Ly-$\alpha$ analysis. Using the CMB similarity, there is a natural quadratic estimator to calculate two-point correlations of the Lyman-$\alpha$ forest (see, for example, \cite{Slosar:2013fi,slosar2011lyman, Busca_2013, white} for details), which can be straightforwardly generalized to the three-point case, as we will discuss later.

Each of the Lyman-$\alpha$ forests, associated to a single QSO, has the ability to sample the underlying density field along a radial segment through the fraction of transmitted light flux $F(\lambda)$ that arrives to Earth. The expression
\begin{equation}
  F(\lambda) = f_q(\lambda)/C_q(\lambda)\,,
  \label{}
\end{equation}
relates the observed flux,
$f_q$, with the un-absorbed one, $C_q$, such that the transmitted fraction $F$ behaves as a biased sampler of the matter
density fluctuations. For the computation of its 2PCF, we define the fluctuations as
\begin{equation}
  \delta_F(\lambda) = \frac{F(\lambda)}{\bar F(\lambda)} - 1\,, 
  \label{delta_field}
\end{equation}
where $\bar F(\lambda)$ is the mean transmitted fraction at the absorber redshift $z$ associated with its correspond wavelength $\lambda$.

\begin{figure}[tbp]
\centering 
\includegraphics[height=7cm]{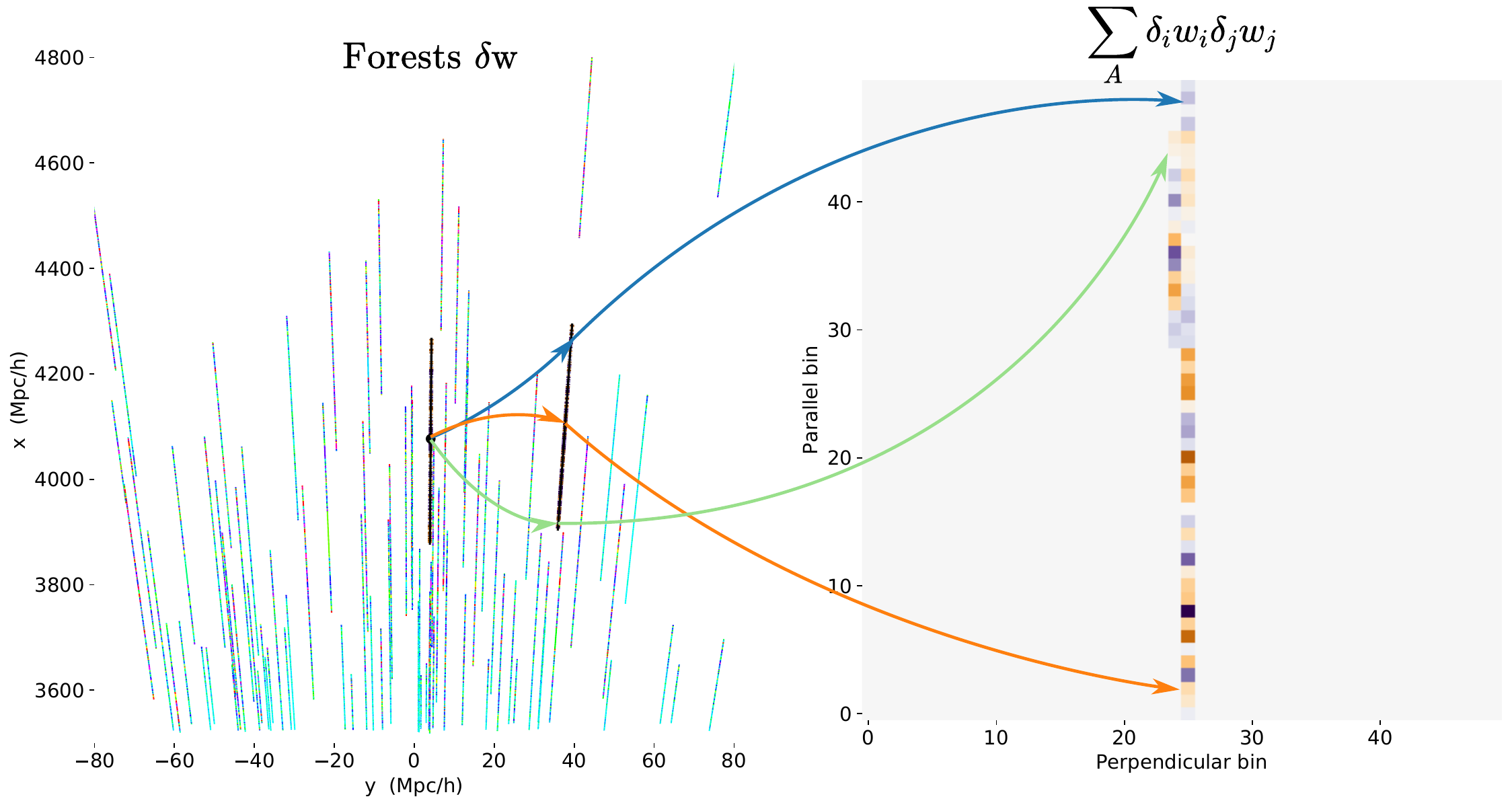}
\caption{\label{fig:pair}  
The forests of a small patch in the sky (left figure) projected in comoving coordinates. Each line points to
Earth which is the origin of the system. The color represents the product $\delta w$. The estimator (\ref{estimator_two})
dictates to multiply all possible pairs of data points between different forests. As an
example, we highlight a pair of forest. A single point in the left forest multiplies to all the right forest.
The results add to specific bins of the grid shown in the figure to the right.}
\end{figure}

\subsection{Two-point correlation estimator}

Using the $\delta_F$-field, an unbiased estimator of the 2PCF can be defined as in ref.~\cite{slosar2011lyman} by
\begin{equation}
  \hat \xi^{(2)}_A = \frac{\sum_A \delta_i w_i \delta_j w_j}{\sum_A w_i w_j} \,,
  \label{estimator_two}
\end{equation}
where the weights, $w_i$, contain different contributions and minimize the total variance of the correlation (for further details on the nature of the weights we refer to \cite{Slosar:2013fi, white, Delubac}). 
The sum is done over all pairs of data points $(i,j)$ from different forests, subject to the condition
that their separation $(r_\bot,r_\parallel)$ in comoving coordinates lies within the bin $A$, which is the rectangle with
$r_\bot \in (r_{\bot A},r_{\bot A} + \Delta r_\bot )$ and equivalent to $r_\parallel$ (see Figure \ref{fig:pair}).
In the computations we present here we use a $4\,\mpch$ bin size and maximal distance of $200\,\mpch$, resulting in 50 bins in each direction or a 2d grid with 2500 entries.

Points within the same forest correlate stronger than between different forests. As a consequence, the former signal was
first measured in \cite{croft1999power}, 12 years before the latter \cite{slosar2011lyman}. We omit the same forest correlations in this work, leading to the so-called three-dimensional correlation, which is widely used for cosmological analyses using the BAO feature.
Our code then computes the correlation function by calculating a histogram of distances between each point in a given forest and points in the other forests, as shown in Figure \ref{fig:pair}.
This algorithm approximately scales as the number of deltas squared, $n_\delta^2=C^2\, N^2$,
where $N$ is the number of quasars (or equivalently skewers) and $C$ is 
the number of deltas sampled per skewer, which peaks around 700 for DESI (DR1 and Y5 mocks), or 200 for eBOSS. For a large number of quasars the fact that we
remove the self-quasar correlations does not really drastically reduce the quadratic scaling. However, to avoid counting beyond the maximal desired separation,
our algorithm only takes forests within a pre-ordered neighborhood defined by this maximum distance, which is easily defined over the HEALPix sky pixelization \cite{healpix}.
This optimization should provide a scaling of the code between $n_b^2$ and the optimal 2pt algorithms (e.g. those using Kd-trees) which approximately scale as $n_b\log(n_b)$. We will see this is the case in the results' section.

\subsection{Distortion matrix}
Due to the estimation of the quasar continuum to extract the $\delta$-field, a 2PCF computation using the estimator (\ref{estimator_two}), $\hat\xi_A$, would be distorted from the true one, $\xi_B$. We consider that the two are related by a linear transformation
\begin{equation}
    \hat\xi_A=\sum_B D_{AB}\xi_B \,,
    \label{}
\end{equation}
and also that the simple model for the continuum estimation (which is encoded in PICCA) is $C_{\rm q}(\lambda) = C(\lambda_{RF}) (a_{q} + b_{q}\log(\lambda)) \label{templatewarpeq} $, with $C(\lambda_{RF})$ a suitable normalization and $(a_q,b_q)$ free parameters.  Under these assumptions, the distortion matrix can be computed as
(see \cite{BOSS:2017fdr} for details) \footnote{Since DR2, the DESI analysis has also included redshift evolution
    considerations, which are not included here \cite{DESI:2025zpo}.}:
\begin{equation}
    D_{AB} = w_a^{-1}\sum_{i,j\in A}w_iw_j\sum_{i',j'\in B}\eta_{ii'}\eta_{jj'}\,,
    \label{distortion}
\end{equation}
with $w_a=\sum_{A} w_iw_j$. The elements $\eta_{ii'}$ are zero except if $i,i'$ are in the same forest, where they follow the expression
\begin{equation}
    \eta_{ii'} = \delta^{K}_{ii'} - \frac{w_{i'}}{\sum_k w_k} -
    \frac{(\Lambda_i-\bar\Lambda)w_{i'}(\Lambda_{i'}-\bar\Lambda)}{\sum_k w_k(\Lambda_{k}-\bar\Lambda)^2}\,,
    \label{}
\end{equation}
with $\delta^{K}$ being the Kronecker delta and $\Lambda = \log \lambda$.
Equation (\ref{distortion}) requires 4 sums over the elements of the forests,
thus the number of computations greatly increases in comparison with the correlation function. In practice $D_{AB}$ is computed
only with a small subset of forest pairs. GPU's can greatly accelerate these computations, as we discuss
later.

\subsection{Three-point correlation estimator}

A straightforward extension for the 2PCF estimator is to obtain an estimator for the three point correlation function (3PCF), by adding a further weighted \textit{delta} product to the expression (\ref{estimator_two}), namely 
\begin{equation}
  \hat \xi^{(3)}_A = \frac{\sum_A \delta_i w_i \delta_j w_j \delta_k w_k}{\sum_A w_i w_j w_k} \,,
  \label{estimator_three}
\end{equation}
where the $i,\ j$ and $k$ run over distinct quasars \footnote{A measurement of three-point correlations on the $\delta$-field over the line of sight can be done in Fourier space, as in the two-point statistics with the 1D power spectrum. This is the so-called 1D bispectrum, which recently has been measured in the eBOSS dataset, showing promising results \cite{arXiv:241009150}.}. 
This extension does not directly implies that the estimator is optimal, since that will depend on the Gaussianity of the estimator's likelihood. However, for mildly non-Gaussian fields, as shown for the CMB \cite{Babich:2005en}, one would expect this estimator to saturate the Cramer-Rao bound, hence to be optimal. That is the case of the Lyman-$\alpha$ forest, which, given the large redshift and low density of the absorption clouds, one would expect the linear evolution of matter perturbations plays a dominant role (see, for example, \cite{Chen:2021rnb}).

\begin{figure}[tbp]
\centering 
\includegraphics[width=5cm]{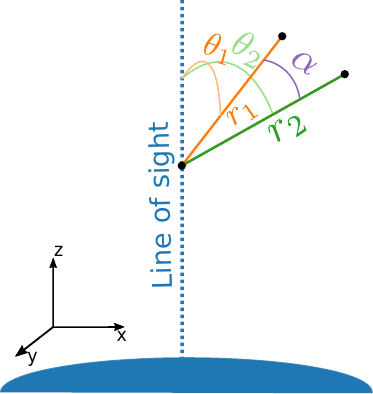}
\caption{\label{fig:triangle} The anisotropic three-point correlation
function depends on the 5 parameters, corresponding to: $r_1$ and $r_2$ the lengths of 
the triangle sides; $\theta_1$ and $\theta_2$ the angle of the triangle sides to the line of sight;
and $\alpha$ the inner angle of the triangle. Note that the figure is in 3D, so in general $\theta_1+\alpha \neq \theta_2$. The same
triangle is recorded in the correlation function depending on which point is taken as the main vertex.}
\end{figure}

In general, the 3PCF depends on 9 parameters, 3 for each delta position. Statistical homogeneity brings that number down to 6 and isotropy reduces it further to only 3. If we consider a triangle defined by one vertex and the displacement vectors $(\boldsymbol{r_1},\boldsymbol{r_2})$, we can choose the parameters as the size of the two sides $(r_1,r_2)$, 
and their opening angle ($\alpha$). The distortions in the redshift space break the full rotational invariance of the correlation functions, reducing this symmetry to rotations about the line of sight. This results in five independent parameters, which can be chosen in several ways. We follow \cite{Slepian:2017lpm} and choose these additional variables as the two angles with respect to the line of sight of the vectors $\boldsymbol{r_1}$ and
$\boldsymbol{r_2}$ (denoted $(\theta_1,\theta_2)$). Figure \ref{fig:triangle} depicts the 5 independent variables we use. In conclusion, the $A$ bins form a 5d hyper-rectangle with sides $(\Delta r, \Delta r, \Delta \theta, \Delta \theta, \Delta \alpha)$, which can be accommodated on a single variable that we call the triangle index.

The number of triplets scales as $n_b$ times the number of pairs from the 2PCF,
hence an extra order of magnitude. Other authors (e.g. \cite{slepian2015computing}) have implemented clever expansion techniques in order to reduce the computations needed to estimate the three-point correlations using multipole expansions. However, in this work, we stick to the brute-force algorithm, relying upon the impressive capacity of GPUs to compute in parallel.

\section{Correlation functions using the Lya2pcf pipeline}\label{3}

The Lya2pcf code computes 2PCF and 3PCF of the Ly-$\alpha$ forest using the estimators (\ref{estimator_two}) and (\ref{estimator_three}).  In order to show three-point measurements and its efficiency, we use two distinct datasets: one from the latest observational data release of the Sloan Digital Sky Survey (SDSS) and the other from the synthetic data which simulate the expected year five catalog of the Dark Energy Spectroscopic Instrument (DESI). Before describing the datasets and our results, we provide a brief description of the code and the computing resources that we use for this work.

\begin{figure*}[tbp]
\centering 
\subfigure[  ]{\label{fig_1}
\includegraphics[width=7cm]{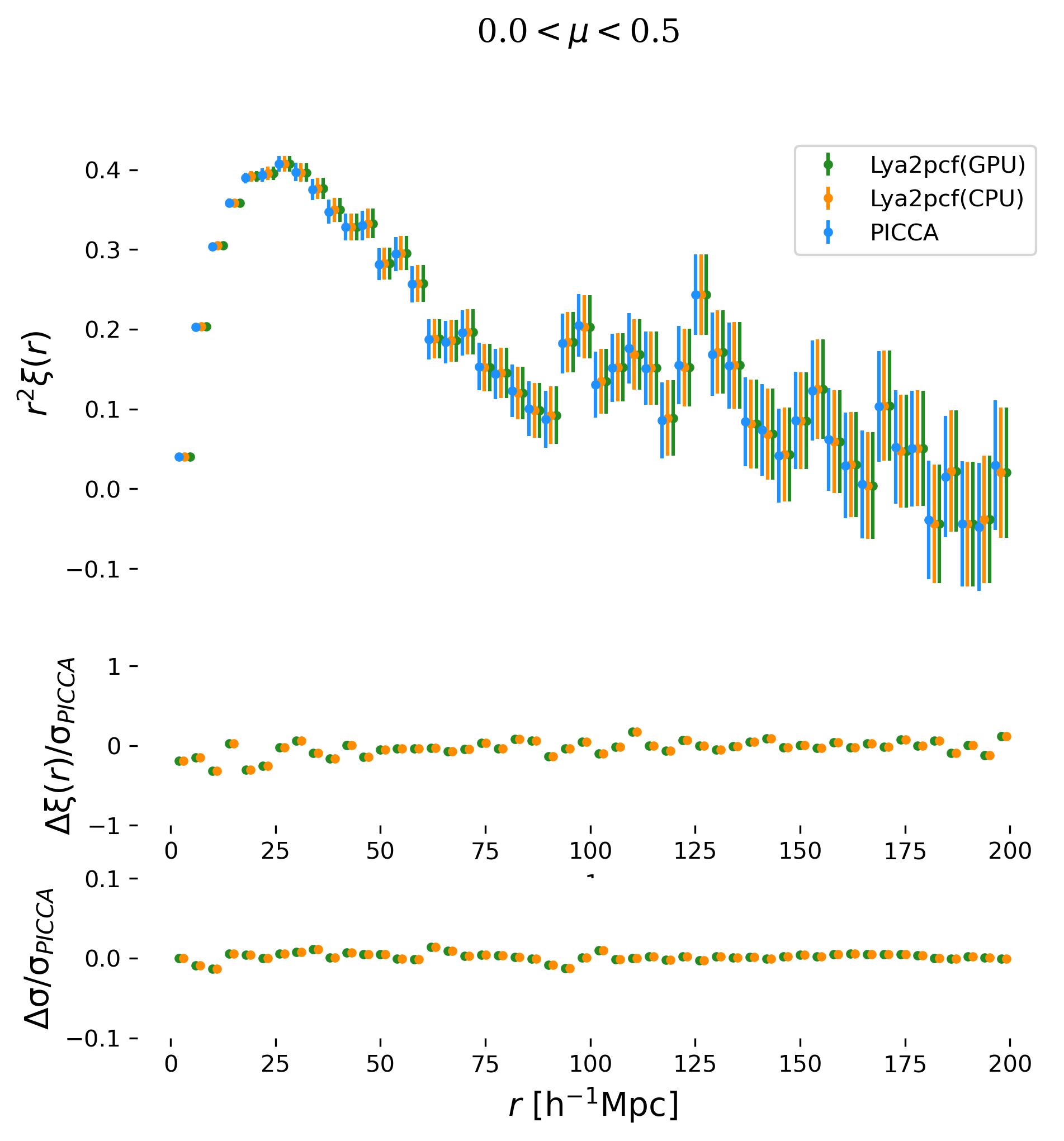}}
\hspace{1em}
\subfigure[  ]{\label{fig_2}
\includegraphics[width=7cm]{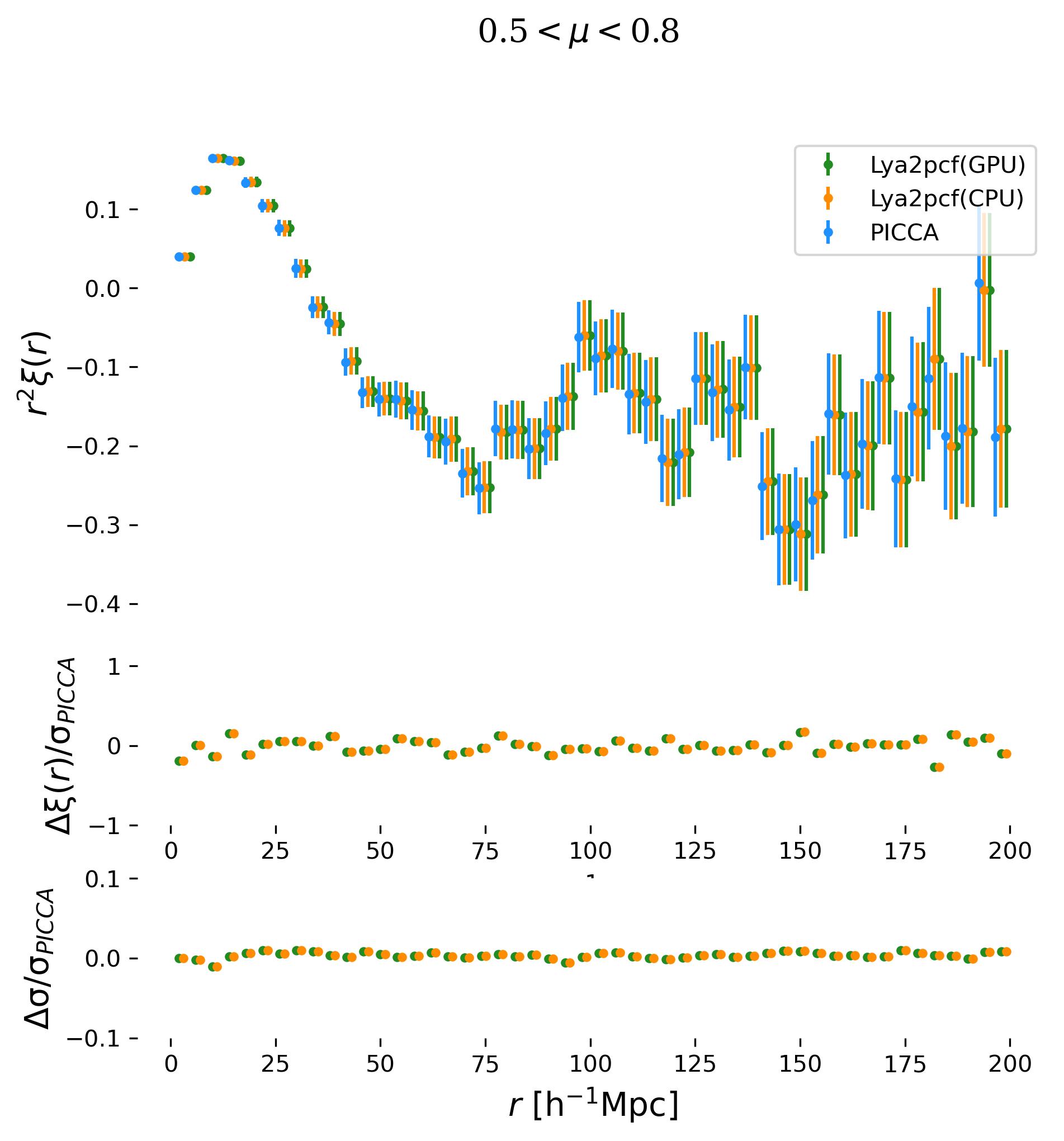}}
\subfigure[]{\label{fig_3}
\includegraphics[width=7cm]{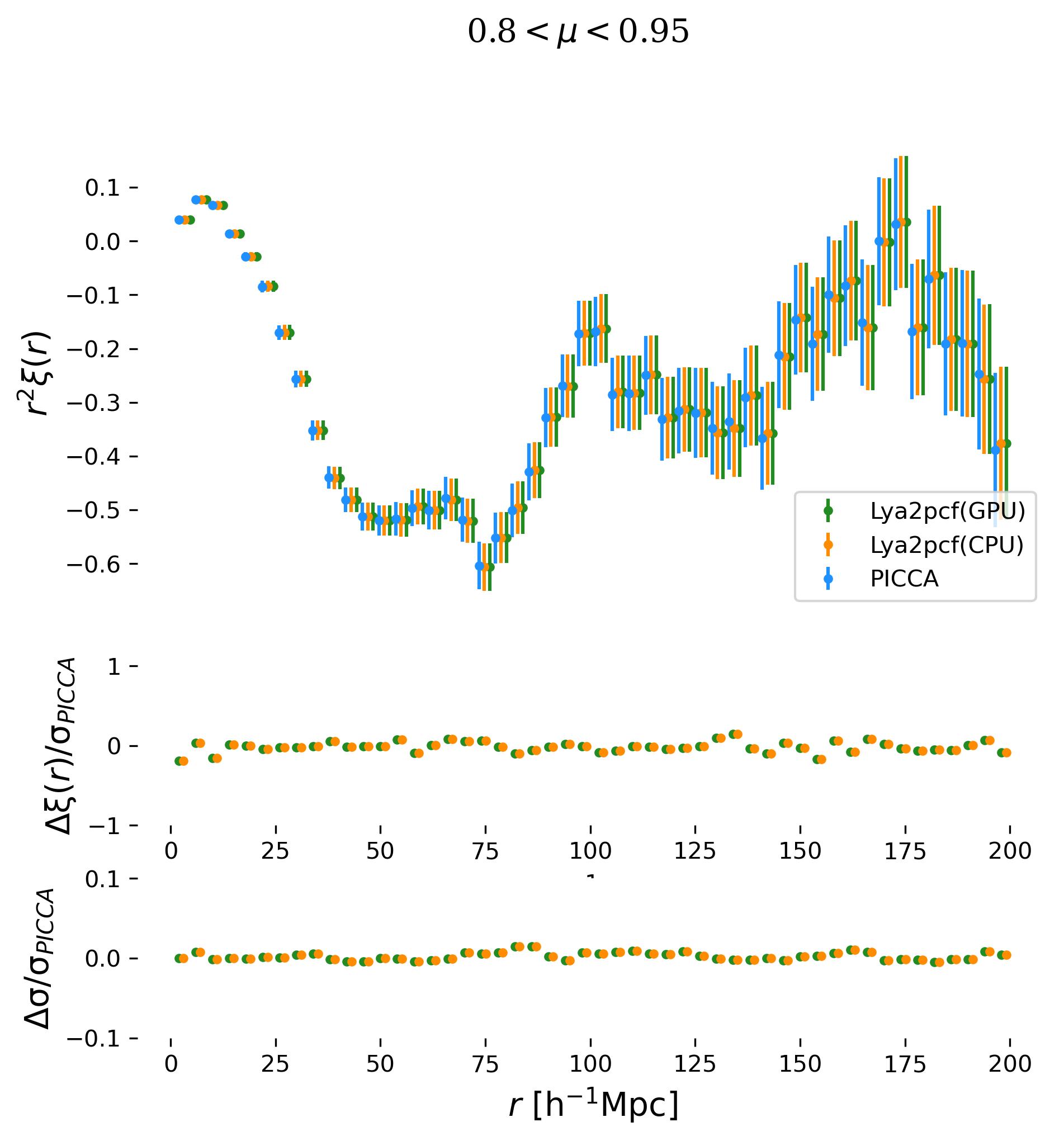}}
\hspace{1em}
\subfigure[]{\label{fig_4}
\includegraphics[width=7cm]{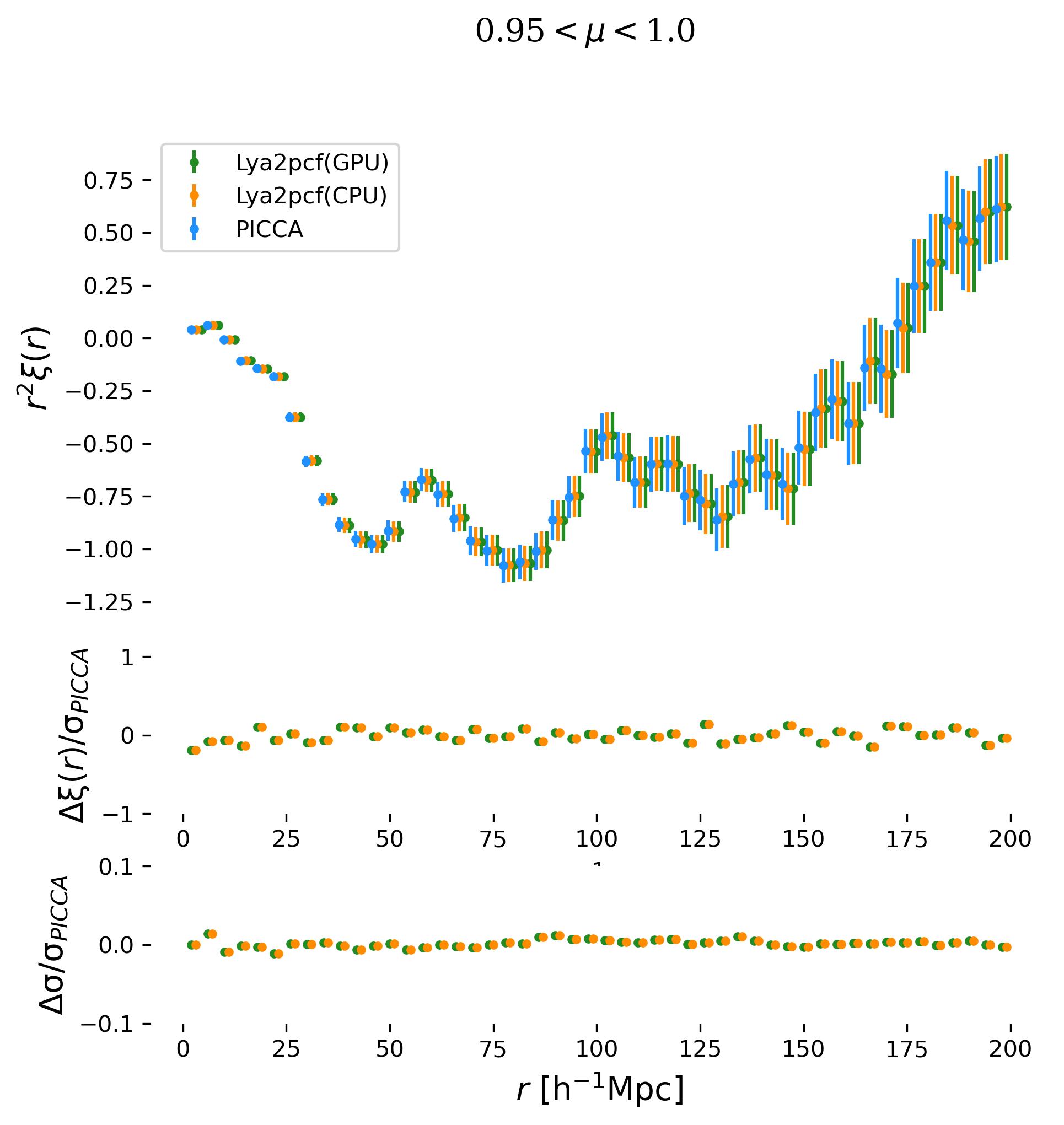}}
\caption{\label{fig:wedges} Wedge plots for the 2PCF for the SDSS DR16 dataset over the angle with respect to the line of sight, $\theta$.
For viewing purposes, we shift the positions in $r$ by a small amount over the different pipeline runs.
Lya2pcf and PICCA clearly show equivalent signals and their corresponding dispersions.}
\end{figure*}

\subsection{Lya2pcf code structure and computational infrastructure}\label{sec:code_infra}

The structure of our code is similar to PICCA. We used Python's high-level capabilities to organize data, along with several useful libraries: Healpy \cite{2005ApJ...622..759G, Zonca2019}, a python wrapper of HEALPix that helps to efficiently find neighboring forests to a given $\delta$ fluctuation; Numba \cite{10.1145/2833157.2833162}, which optimizes the computationally intensive functions;
MPI4Py \cite{9965751}, which allows us to efficiently distribute the work load between different nodes (for both architectures, CPUs and GPUs); and PyCUDA \cite{2009arXiv0911.3456K} to program directly CUDA kernels that will be executed in the GPU, being this the only non-python part of the pipeline. In particular, in order to compute the sums depicted in Figure \ref{fig:pair}, the Just-In-Time Numba compiler was used for the CPU computations while CUDA kernels where written for the GPU case and linked to the Python code with PyCUDA. 
When using only CPUs, a function computes all the sums for a given pair of forests; meanwhile in the case
of GPUs, each kernel computes the histogram for all the pairs between a single forest and all of its neighbors.
In terms of efficient algorithms, for the distortion matrix, which takes the largest computation time in the 2PCF pipeline, we only do it over GPUs and using PyCUDA.

Our code is adapted to read the delta files used by SDSS and DESI. Which consists on a set of FITS files, one for each HEALPix pixel.
In the SDSS format, each skewer is stored in a Header/Data Unit
with its position (right ascension and declination) and a table of variable lenght
containing its deltas, weights and wavelengths.  In contrast, DESI uses
a single table for all the deltas in the pixel and one for all the weights, requiring the
skewers to be of the same length with
\code{Nan} placeholders for missing wavelengths.
The code's output mimics PICCA generating a FITS file with the correlation, covariance and distortion matrix, which can be used by the code Vega\footnote{\url{https://github.com/topics/lyman-alpha}} to fit different model parameters, particularly to measure the BAO scale.
These three outputs are also saved separately as NumPy arrays for plotting in a provided Jupyter notebook.

We use three different computing architectures for this study. To fairly compare Lya2pcf and PICCA, we run both codes under the same CPU infrastructure: a server with two Intel Xeon 8 Cores E5 2.1 GHz processors. The GPU performance for the 2PCF computation of the largest data set is studied on two NVIDIA A100 Tensor Core GPUs with 80 GB of memory\footnote{Additional specifications in \href{https://www.nvidia.com/content/dam/en-zz/Solutions/Data-Center/a100/pdf/nvidia-a100-datasheet-us-nvidia-1758950-r4-web.pdf}{A100}.} in the NERSC system. To measure the anisotropic 3PCF as well as the time scaling relations of the 2PCF we decided to use a lower performance GPU card, GeForce RTX 2070\footnote{Further details in \href{https://www.nvidia.com/content/geforce-gtx/GEFORCE_RTX_2070_User_Guide.pdf}{GeForce RTX 2070}.}, which more accurately exemplifies the average running cost on a personal computer. 

We use the \code{time} routine in Linux to measure the real execution time and average over 5 executions for each use case. The version of PICCA that we use for time comparisons is 9.0.4, which is available at \url{https://github.com/igmhub/picca/}. 

\subsection{Datasets}

The 3PCF study is centered on the observational spectra of quasars in the redsfhit range $2.1<z<4$, contained in the sixteenth data release (DR16) of the Sloan Digital Sky Survey (SDSS) \cite{2020ApJS2508}. The sample consists of 210,005 Ly-$\alpha$ forests with a mean signal-to-noise of 2.56, whose $\delta$ field has been calculated using a particular version of PICCA\footnote{The used PICCA version is in the repository \url{https://github.com/igmhub/picca/releases/tag/v4.alpha}.}. We use directly this $\delta$-field catalog, which has been made public by the SDSS Collaboration\footnote{Further details on the data can found in \url{https:/www.sdss4.org/dr17/spectro/lyman-alpha-forest/}.} in the repository \href{https:/data.sdss.org/sas/dr16/eboss/lya/}{data.sdss.org}. We consider only the Lyman-alpha forest region, which is split into HEALPix pixels (with an nside=8). We do not do further cuts in the data, which is well described in \cite{des2020completed}, where the reader can find details of the weights arising from the instrument's noise and the Large Scale Structure variance.

We also use a set of synthetic spectra provided by the DESI Lyman-$\alpha$ working group, which has a realistic angular, redshift, and magnitude distribution of quasars of what could be expected by the end of fifth year of operation. This synthetic spectra is built on approximate simulations of the large-scale density field, with the LyaCoLoRe code, from which one can extract density skewers \cite{2020JCAP03068} that are revisited afterwards with quasar properties, like the quasar continuum, emission lines, redshift errors, etc. The details of how these synthetic spectra were generated are in \cite{2025JCAP01141}; in particular, Section 5 explains how a very similar dataset was used to forecast the constraining power of the full DESI survey. Other uses of similar datasets can be consulted in \cite{Youles:2022pgf, DESI:2024txa}. 
Our particular interest to use these simulated data is to compare the performance, in terms of computing time, between our implementation and the PICCA code. In this sense, the most important characteristic is the size of the simulated dataset, which corresponds to 1,015,588 simulated spectra. Nevertheless, as we show in the results' section, the two codes lead to a similar signal for this DESI Y5 dataset, as well as for the eBOSS data. Table  \ref{tab:forecast} summarizes the characteristics of both Ly-$\alpha$ catalogs: SDSS DR16 and a DESI Y5 mock.

\begin{table}[tbp]
\centering
\begin{tabular}{|c|c c c|}
\hline
Data sets & $N_0$ forests & Area & $\Delta x$\\
\hline
SDSS DR16            & 210,005 & 10,563.35 deg\textsuperscript{2} & 3.5 Mpc/$h$ \\
DESI Y5 mocks            & 1,015,588 & 15,765.29 deg\textsuperscript{2} & 0.6 Mpc/$h$ \\
\hline
\end{tabular}
\caption{\label{tab:forecast} Data information in SDSS DR16 and the DESI Y5 mocks, showing the number of forests $N_0$, area and the mean separation between deltas in comoving distance. The number of forests in the DESI Y5 mocks is approximately 4.8 times larger than in SDSS DR16. }
\end{table}

\subsection{Results: measured signal and code performance}
For the two-point statistics, our code gives results similar to those obtained by the widely used code PICCA. The difference in the measured signals with both codes, as well as their corresponding dispersions, is consistent with each other, as seen in Figure \ref{fig:wedges}. Actually, variations of our code using CPUs and GPUs are of the same order as those with respect to PICCA. Consistent results are also obtained for the distortion matrix. 

\begin{table}[tbp]
\centering
\begin{tabular}{|c c c c c|}
\hline
 Pipeline & Computation & Dataset & Hardware used &  Time \\
  \hline 
  Picca & 2PCF & Desi Y5 mocks & $2\times 8$ core CPUs & 37h \\
  Picca & Distortion matrix & Desi Y5 mocks & $2\times 8$ core CPUs  & 205h \\
  Lya2pcf CPU & 2PCF & Desi Y5 mocks & $2\times 8$ core CPUs  & 24.1h \\
  Lya2pcf GPU& 2PCF & Desi Y5 mocks & $2\times$A100 GPUs & 4.3h \\
  Lya2pcf GPU& Distortion matrix & Desi Y5 mocks & $2\times$A100 GPUs & 6.4h \\
  Lya2pcf GPU& Anisotropic 3PCF & SDSS DR16 & Geforce GPU & 1.7h \\
\hline
\end{tabular}
\caption{\label{tab:cpu_equiv} Time comparison for the largest datasets on each computation category, which includes two and three point correlations and the distortion matrix for the two-point case. PICCA, when calculating the 2PCF and the distortion matrix, takes 22.6 times longer than Lya2pcf using GPUs. 
Our pipeline on GPUs performs better, most notably for larger datasets (see Figure \ref{fig:Nforests}).
Real execution times are shown for correlations up to $\rmax=200\mpch$ for the 2PCF and $40\mpch$ for the 3PCF. An overhead time of 2.3h was added to the execution times of the Lya2pcf 2PCF coming from a preparatory step where the \code{fits} file information is extracted (See Appendix \ref{A}). Specifications of GPU and CPU architectures are explained in section \ref{sec:code_infra}.}
\end{table}

However, in terms of computational costs, Lya2pcf and PICCA perform differently. As shown in Table \ref{tab:cpu_equiv}, PICCA takes longer to calculate two-point correlations when both codes are run under the same CPU architecture, with Lya2pcf taking 65\% of the PICCA running time. Furthermore, the GPUs ability to make highly parallel computations fits particularly well for the problem at hand,
making Lya2pcf faster on GPUs than on CPUs, especially for higher work loads such as with the DESI year five catalogs. This great improvement is better appreciated when calculating the distortion matrix. For example, Table \ref{tab:cpu_equiv} 
shows only a small improvement for the two-point correlation function when using CPUs or GPUs,
where the overhead time needed to upload the $\delta$'s information to the GPU compensates the gain in computation speed.
The total computation time of the distortion matrix using the GPU shows a considerable gain, which can be appreciated from the quotient of times taken by the 2PCF and distortion matrix (DM) computations using PICCA under CPUs ($t_{DM}/t_{2PCF}=5.54$) or Lya2pcf under GPUs ($t_{DM}/t_{2PCF}=3.2$).

\begin{figure}[ht]
\centering 
\includegraphics[width=11cm]{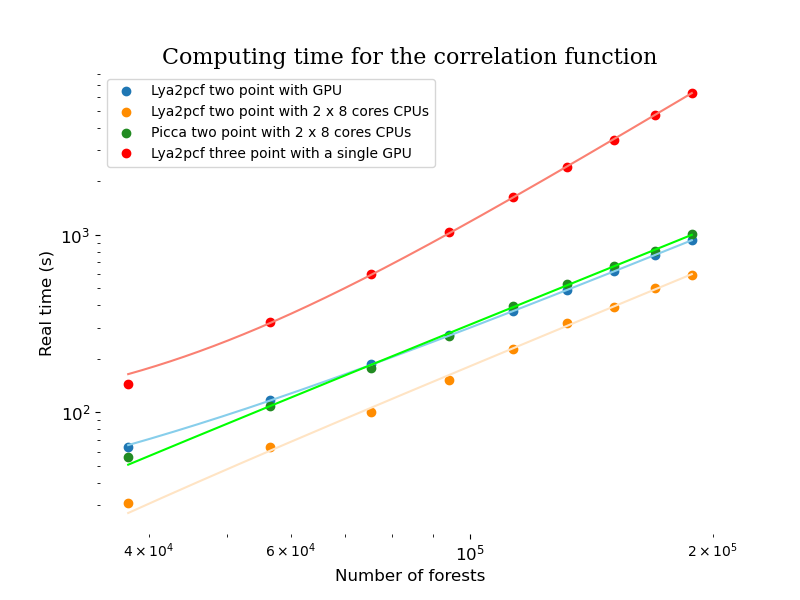}
\centering 
\includegraphics[width=11cm]{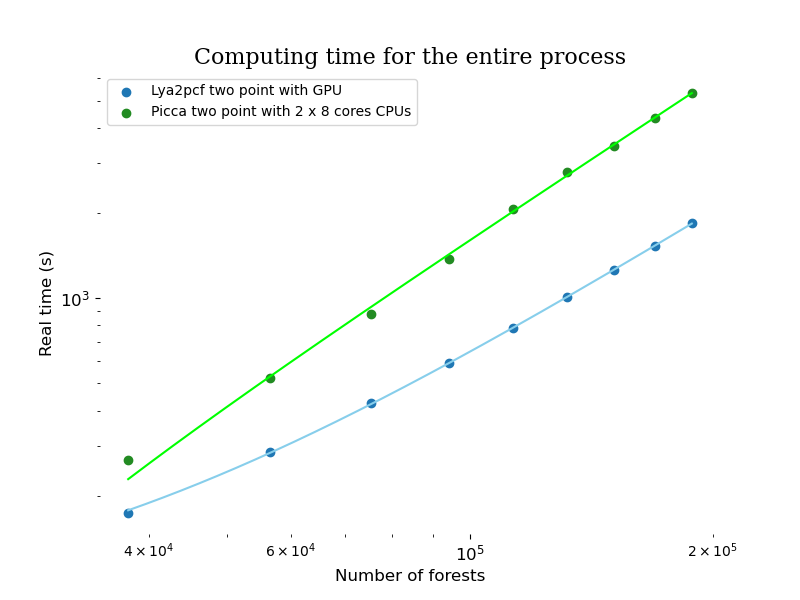}
\caption{\label{fig:Nforests}
Execution time for the computation of the correlation function and distortion matrix for different total number of forests using data from SDSS DR16. The GPU times were obtained using a single GeForce RTX 2070 graphics card.
For the 2PCF we used $\rmax=200\mpch$ while only $40\mpch$ for the 3PCF.
The continuous curves are obtained by fitting a function $t = t_0 + k (N/N_0)^b$ with the parameters written in Table \ref{tab:Nforests}. We can see that the computing time for the correlation function does not differ significantly between Lya2pcf-GPU and PICCA executions. However, when considering the computing time of the distortion matrix, we see that the time between the executions of Lya2pcf and PICCA differs significantly (see Table \ref{tab:cpu_equiv}).}
\end{figure}

For a more quantitative understanding of computational costs for large galaxy surveys, we test the running time of PICCA and Lya2pcf as a function of the number of forests, using different subsets of the SDSS DR16 data. Results are shown in Figure \ref{fig:Nforests}.
We adopt a model which assumes that the time consists on the addition of an initialization time $t_0$ plus the main computation, which we parametrised as a power law function of the number of forests; namely
\begin{equation}
  t(N) = t_0 + k \left( \frac{N}{N_0} \right)^b \,,
  \label{power}
\end{equation}
where we have decided to normalise the expression by $N_0$ factor, corresponding to the total number of forests that we take from SDSS DR16 ($N_0=210,005$).
Table \ref{tab:Nforests} shows the fitted parameters for the different types of execution. We see that the scaling exponent for the 2PCF is close to two for both PICCA and Lya2pcf, while it approaches three for the 3PCF. The exact values of two and three are expected for an estimator that does not have an efficient tree-searching structure for near neighbors. However, PICCA and Lya2pcf use Healpix pixels to identify which forests are close to each other. An optimised tree search for the neighboring forests would further improve these scaling relations (see for similar approach in weak lensing on 2d \cite{Arvizu:2024rlt}), an improvement line that will pursue in the future.

\begin{table}[tbp]
\centering
\begin{tabular}{|c|c c c|}
\hline
Two-point correlation (2PCF) & $t_0$(minutes) & $k$(minutes) & $b$\\
\hline
PICCA using a CPU            & 0.0 & 20.3 & 1.8 \\
Lya2pcf using a CPU            & 0.0 & 12.3 & 1.9 \\
Lya2pcf using a GPU            & 0.3 & 18.6 & 1.9 \\
\hline
Two-point correlation (2PCF) $+$ Distortion matrix & & & \\
\hline
PICCA using CPU & -0.6 & 109.2 & 1.9 \\
Lya2pcf using GPU & 1.3 & 35.5 & 1.8 \\
\hline
Anisotropic three-point correlation (3PCF$_{aniso}$) & & & \\
\hline
Lya2pcf using a GPU& 1.5 & 138.8 & 2.7 \\
\hline

\end{tabular}
\caption{\label{tab:Nforests} Best fit parameters for the computing time as function of the number of forests.
We fitted the function $t=t_0+k(N/N_0)^b$ with
$N_0=210,005$ (total number of SDSS DR16 forests). Real execution times are shown for correlations up to $\rmax=200\mpch$ for the 2PCF and $40\mpch$ for the 3PCF.  The GPU times were obtained using a single GeForce RTX 2070 graphics card.
Time-scalings, given by the $b$ exponent, show a mild dependence on the architecture, with faster results on GPUs. Present day computer infrastructures can reasonably measure the anisotropic 3PCF for the larger datasets in the stage IV spectroscopic surveys.}
\end{table}

\begin{figure}[tbp]
\centering 
\includegraphics[width=\textwidth]{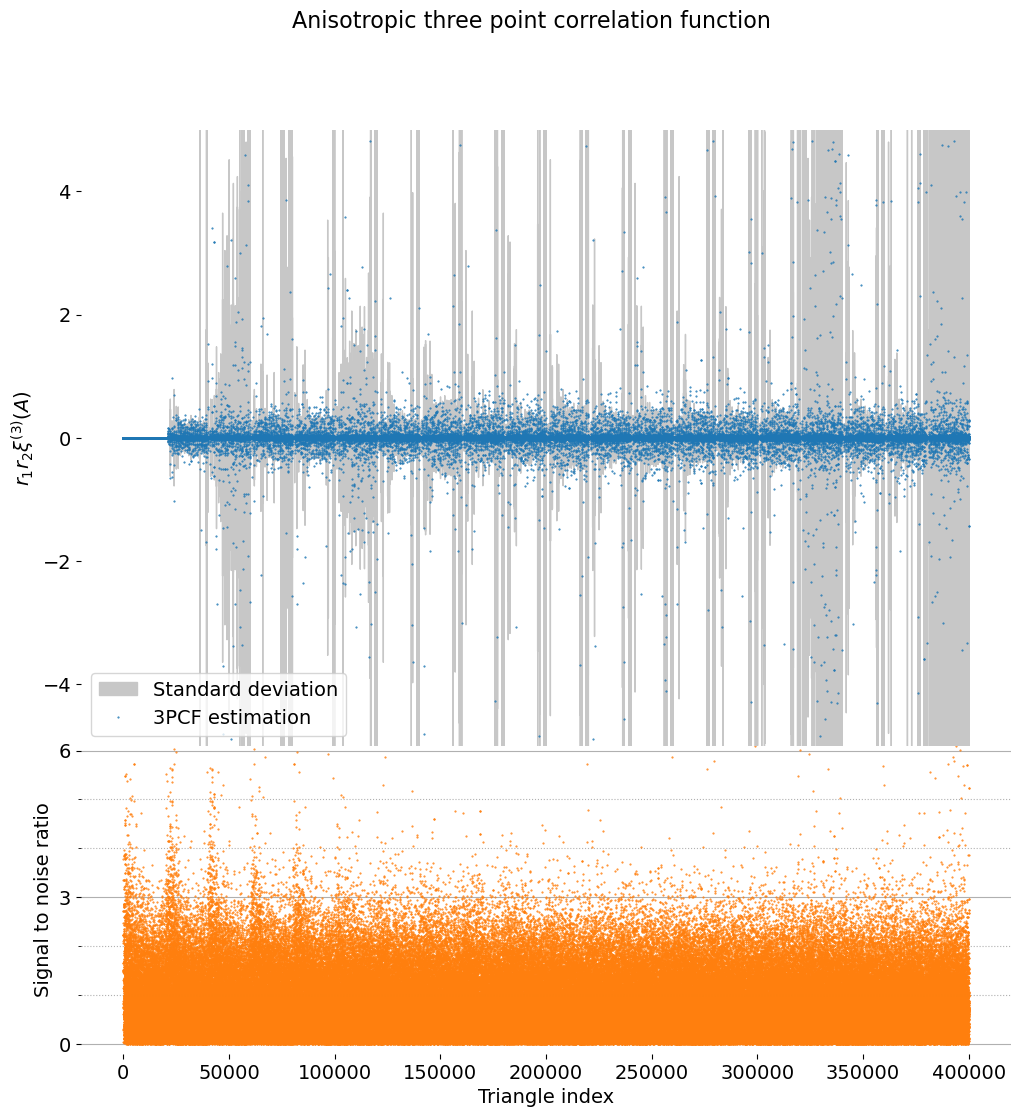}
\caption{\label{fig:3pcf} First measurement of the anisotropic Ly-$\alpha$ 3PCF of the SDSS DR16 dataset. We show triplet correlations as a function of the triangle index $A$, which flattens with triangle variables $(r_1,r_2,\cos(\alpha),\theta_1,\theta_2)$, with
$r_i$ from 0 to 80 $\mpch$ divided in 20 bins,
$\theta_i$ from 0 to $\pi$, in 10 bins; and $\cos(\alpha)$ from $-1$ to 1, in 20 bins.
The standard deviation is obtained by subsampling the forests' data in Healpix pixels.}
\end{figure}

One of the main reasons to write Lya2pcf is to be able to compute the 3PCF. Although previous analyses of the three point statistics in the Lyman-$\alpha$ exist in the literature, they mostly have been devoted to the one dimensional bispectrum \cite{Viel:2003fp, Mandelbaum, arXiv:241009150}. For the 3PCF in real space, among the relevant work is that of \cite{Tie:2019tpi}, where it was predicted the isotropic 3PCF signal for particular triangle configurations using synthetic data, with the aim of understanding UV background fluctuations. The authors suggest a signal-to-noise (S/N) of around 9 for the eBOSS sample. The goal in this work is complementary to \cite{Tie:2019tpi}, providing the first measurement of the full anisotropic 3PCF over observational Lyman-$\alpha$ forests. Measuring this 3PCF signal is the first step towards using the 3PCF estimator (\ref{estimator_three}) to constrain cosmology and the properties of the IGM. As it can seen in Figure \ref{fig:3pcf}, we have promising results in the DR16 data sample, where we show the anisotropic 3PCF signal and its dispersion using subsamples of the data. All triangle configurations are shown, with their five parameters accommodated in a single (triangle) index. The S/N of around 22.5\% (1.2\%) of the binned configurations is above one (three), where the small-scale isosceles triangles are particularly dominant. Notice that the anisotropic signal has less restrictive power than the isotropic case, hence its smaller S/N than the predicted value of 9 of \cite{Tie:2019tpi}. Actually, while averaging over both $\theta_i$ angles would result in the isotropic signal, its dispersion, re-scaled by the appropriate power of the number of bins, would imply a S/N$\sim 9$ for triangles where both sides are under 20 Mpc/h.

\section{Discussion and future prospects}\label{4}

It is well known that higher-order statistics can be used in cosmology to extract some of the non-linear information of the matter distribution. Part of this information is unique to each n-point statistics, thus not contained in the traditional two-point correlations. 

Moreover, galaxy surveys nowadays have the right capabilities to fully explore the high-redshift regime, where these non-linearities in the matter distribution are still mild. In particular, the spectra of high-redsfhit quasars have absorption \textit{forests} that trace the matter distribution between these sources and us along the line of sight. 
Although one expects that a linear model of the matter distribution at those redshifts is accurate enough to describe the forest correlations, with the precise data of present and future galaxy surveys, we are in a position to characterize deviations from the Gaussian field. A first approach for this characterization of the non-linear behavior comes from studying corrections to the modeling of the 2PCF, which has been included in the modeling of the traditional Lyman-$\alpha$ analysis (see, for example, the latest DESI DR2 Ly-$\alpha$ analysis). More recently an EFT description (e.g. \cite{Hadzhiyska:2025cvk}) offers new insights on this nonlinearities. A complementary approach is to measure higher-order statistics, in particular the next order which is the three-point correlation function. This work represents a first step towards this perspective. 

In this work, we measure a non-trivial three-point correlation function (3PCF) using 210,005 Lyman-$\alpha$ forests in the SDSS DR16 sample. In particular, we explore the signal of anisotropic triangle configurations, which can be described by five parameters. We use two sides of the triangle, the opening angle between those sides, and two additional angles characterizing the orientation with respect to the line of sight. Our binned measurement has all possible angles and scales up to 80 Mpc/h. The calculation time to correlate triplets scales with an additional factor of $N$ with respect to the 2PCF, where $N$ is the number of data points. As a result, we rely on GPUs for this computation, which can be done in a reasonable amount of time using commercial GPU cards, such as the Geforce RTX 2070. 

The signal-to-noise of the anisotropic Ly-$\alpha$ 3PCF in the SDSS DR16 sample is above one for almost a quarter of triangle configurations, exhibiting the non-linear information of the forest fluctuations and its viability to be used to constraint cosmological parameters, as well as information of the intergalactic medium. During the completion of this work, the DESI collaboration delivered its First Data Release (DR1) \cite{DESI:2025qqu}, whose Ly-$\alpha$ forest catalog should have a stronger 3PCF signal. However, as shown in Figure \ref{fig:3pcf}, the three-point signal is rather complicated, therefore, in order to make further use of 3PCF measurement with DESI, an analytic description of the signal would be necessary. This full-shape modeling can be done using the recent EFT ideas of \cite{Ivanov:2023yla}, but we leave this line of research to future studies, where we will focus on the most recent DESI data. 

The exploration of higher-order statistics in the fluctuations of the Ly-$\alpha$ forests led us to build a dedicated pipeline, whose starting point is the delta field, or in other words, the forest fluctuations with respect to an estimated quasar continuum. We used the delta-extraction of the public PICCA to obtain this delta field, whose continuum model introduces a bias in the correlation functions. An ad-hoc distortion matrix restores most of the unbiased 2PCF, and an equivalent prescription would have to be extended to the three-point estimator. 

In the spirit of an alternative code to the widely adopted tool PICCA for computing two-point Ly-$\alpha$ auto-correlations, we also include a 2PCF module in our pipeline analysis, with the estimation of its corresponding distortion matrix. The use of GPUs for this matrix is, again, more efficient, providing an optimal framework for analyzing large datasets. In the computer architectures used here, we find an improvement of up to $\sim 30$ times faster than PICCA, when calculating the Ly-$\alpha$ autocorrelation (with the distortion matrix) of the expected year-5 DESI Ly-$\alpha$ data. Future improvements in our pipeline will include one-dimensional correlations, as well as cross-correlations between forests and quasars. Moreover, even though the scaling of our algorithm is better than the brute-force approach due to the Healpix pre-ordering of skewers, there is still room for refinements using tree algorithms to speed the search for neighboring pairs, in both the 2PCF and 3PCF.

Finally, we believe that it is important to have an alternative pipeline to validate the standard results obtained by PICCA for two-point correlations of the Ly-$\alpha$ forest in present-day surveys such as DESI. Therefore, our pipeline has already been adapted to use the data currently obtained and released by DESI. In addition to running Lya2pcf with the DESI Y5 mocks, our pipeline was also tested with DESI's public releases: Early Data Release (EDR) \cite{2024AJ16858} and Data Release 1 (DR1) \cite{arXiv:2503.14745}. The correlation function and distortion matrix using Lya2pcf are consistent with those obtained using PICCA for these public DESI datasets, with the benefit of reducing computational times when running on GPUs.

\subsection{Code availability}
A first version of the Lya2pcf code is available at \url{https://github.com/Rafael9104/lya2pcf}. At the time of writing, this free public version computes the two-point auto-correlation function, covariance matrices and the associated distortion matrix. It can take data using SDSS or DESI data formats. A future release will incorporate the three-point correlation function, as well as other improvements. For details on how to install and run Lya2pcf, we refer the reader to Appendix A, as well as to the README file in the code's repository.

\acknowledgments
The authors acknowledge the initial guidance of
Hélion du Mas and Andreu Font in this project; as well as the helpful discussions with Julian E. Bautista, Debopam Som and Suk Sien; and Julio Clemente and Octavio Valenzuela for guidance with the computational optimizations. We also thank the DCI-UG DataLab for computational resources and the DESI Lyman-$\alpha$ working group for providing access to one of their Y5 mocks to test the performance of our pipeline, in particular to Hiram K. Herrera-Alcantar. RGB was supported by SECIHTI grant 773222. GN acknowledges the support of SECIHTI (grant ``Ciencia Básica y de Frontera'' No. CBF2023-2024-162), DAIP-UG, and the Instituto Avanzado de Cosmologia. 

\appendix

\section{Installing and running Lya2pcf}\label{A}
To use Lya2pcf, it is necessary to install the Python libraries found in the
\code{README} file, or using \code{requirements.txt} with the command
\begin{itemize}
    \item \code{conda install --yes --file requirements.txt}
\end{itemize} 
To calculate Lyman-$\alpha$ 3D correlations, one needs the $\delta$-field, which can be obtained using Picca or any other extraction method because, as
currently developed, the Lya2pcf code does not offer an alternative to calculate these fluctuations. Notice that a useful function for converting common FITS file to their numpy counterparts is
\begin{itemize}
\item \code{python delta\textunderscore reader.py --delta-dir DELTA\textunderscore DIR}
\end{itemize}
Once the $\delta$-field is provided, one can calculate 2PCFs using CPUs or GPUs, running the following command
\begin{itemize}
    \item \code{python 2pla.py (--cpu | --gpu)}
\end{itemize}
The distortion matrix can be obtained by running
\begin{itemize}
    \item \code{python distortion.py}
\end{itemize}
Note that this part of the calculation can only be performed over GPUs.
The 2PCF sums allow to construct the covariance and correlation using
\begin{itemize}
\item \code{python post\textunderscore processing.py}
\end{itemize}
This last step also outputs a \code{fits.gz} file that can be used with the fitter Vega.
Further specifications can be found in the \code{README} file in the code repository.

\bibliographystyle{jhep}
\bibliography{Bibliography}

\end{document}